\documentclass[aps,pre,amsfonts,floatfix, onecolumn,showkeys]{revtex4-1}
\usepackage{amssymb}
\usepackage{amsmath}
\usepackage{hyperref}
\usepackage{graphicx}
\usepackage{dcolumn}
\usepackage{enumerate}
\usepackage{times}
\usepackage{graphicx}
\usepackage{multirow}
\usepackage{times}
\usepackage{lineno}
\usepackage[english]{babel}
\usepackage{typearea} 
\usepackage{color}
\usepackage{epstopdf}

\begin{document}

\title{
Mutualism and evolutionary multiplayer games: Revisiting the Red King
}

\author{Chaitanya S. Gokhale}
 \email{gokhale@evolbio.mpg.de}
\author{Arne Traulsen}%
\affiliation{%
Research Group for Evolutionary Theory,\\
Max-Planck-Institute for Evolutionary Biology,\\
August-Thienemann-Stra{\ss}e 2, 24306 Pl\"{o}n, Germany}%

\begin{abstract}
Coevolution of two species is typically thought to favour the evolution of faster evolutionary rates helping a species keep ahead in the Red Queen race,
where `it takes all the running you can do to stay where you are'.
In contrast, if species are in a mutualistic relationship, it was proposed that the Red King effect may act, where it can be beneficial to evolve slower than the mutualistic species.
The Red King hypothesis proposes that the species which evolves slower can gain a larger share of the benefits.
However, the interactions between the two species may involve multiple individuals.
To analyse such a situation, we resort to 
evolutionary multiplayer games. 
Even in situations where evolving slower is beneficial in a two-player setting,
faster evolution may be favoured in a multiplayer setting.
The underlying features of multiplayer games can be crucial for the distribution of benefits.
They also suggest a link between the evolution of the rate of evolution and group size.
\end{abstract}

\keywords{
mutualism, evolutionary game theory, multiple players, rate of evolution}

\maketitle



\section{Introduction}

Mutualistic relationships, interspecific interactions that benefit both species, have been empirically studied for many years \cite{boucher:book:1985,hinton:PTENHS:1951,wilson:AmNat:1983,bronstein:QRB:1994,pierce:ARE:2002,kiers:Nature:2003,bshary:book:2003} and also a considerable body of theory has been put forth trying to explain the evolution and maintenance of such relationships \cite{poulin:JTB:1995,doebeli:PNAS:1998,noe:book:2001,johnstone:ECL:2002,bergstrom:PNAS:2003,hoeksema:AmNat:2003,akcay:PRSB:2007,bshary:Nature:2008}.
Many of these studies utilize evolutionary game theory for developing the models.
The interactions in these models are usually pairwise.
A representative of each species is chosen and the outcome of the interactions between these representatives 
determines the evolutionary dynamics within each of the two species.
However, in many cases interactions between species cannot be reduced to such pairwise encounters \cite{stadler:book:2008}.

For example, in the interaction between ants and aphids or butterfly larvae \cite{kunkel:BZB:1973,pierce:BES:1987,hoelldobler:book:1990} many ants tend to these soft bodied creatures, providing them with shelter and protection from predation and parasites in exchange for honeydew, a rich source of food for the ants \cite{hill:OEC:1989,stadler:book:2008}.
This is not a one to one interaction between a larva and an ant, but rather a one to many interaction from the perspective of the larva.
In this manuscript we focus on this kind of -- possibly -- many to many interactions between two mutualistic species.

To analyse how the benefits are shared between the two mutualistic species, we make use of evolutionary game theory \cite{weibull:book:1995,hofbauer:JMB:1996,hofbauer:book:1998}.
Following Bergstrom \& Lachmann \cite{bergstrom:PNAS:2003}, we  analyze the interactions between two species with a twist to the standard formulation.
The two interacting species can have different evolutionary rates.
In coevolutionary arms races, where species are locked in antagonistic relationships such as host-parasite interactions, we observe the Red Queen effect \cite{vanValen:EvoTheo:1973}.
In these cases a higher rate of evolution will be beneficial, as it would help the parasite infecting a host or a host escaping from a parasite.
However, in mutualistic relationships a slower rate of evolution was predicted to be more favourable \cite{bergstrom:PNAS:2003}.
Hence, it is possible that the type of relationship between two species affects the evolution of the rate of evolution.
As mentioned in \cite{bergstrom:PNAS:2003} and \cite{dawkins:PRSB:1979} the different evolutionary rates could be due to a multitude of factors ranging from different population sizes to the differing amount of segregating genetic variance.
The implications of a difference in evolutionary rates are not limited to mutualism and antagonistic relationships.
Epidemiological modeling and data have shown correlations between the rate of epidemic spreading and the evolutionary rate of the spreading pathogen \cite{berry:JV:2007,zehender:VIR:2008}.

We first recall the mutualistic relationship between two species in a two player game, as proposed by Bergstrom and Lachmann.
Then, we increase the number of players.
Note that we do not increase the number of interacting species \cite{mack:OIKOS:2008,damore:Evolution:2011}, but rather the number of interacting individuals between two species (also see \cite{wang:JRSI:2011}).
We include asymmetry in evolutionary rates and discuss its effect both in two player and in multiplayer games.
We find that when in a two player setting it is beneficial to evolve at a slower rate, it can be detrimental in a multiplayer game.

\section{Model and Results}

To set the stage we first recapitulate the two player model presented in \cite{bergstrom:PNAS:2003} albeit with different notation.
They considered two species each with two strategies, \textit{``Generous"} and \textit{``Selfish"}.
Each species is better off being \textit{``Selfish"} as long as the other one is \textit{``Generous"}.
If both are \textit{``Selfish"}, then no mutualistic benefit is generated and hence in that case it is better to be \textit{``Generous"}.

Under these assumptions, the payoff matrices for the interactions describing the interactions of each type with one member of the other species are
\begin{equation}\label{}
\begin{array}{cccc}
& & \multicolumn{2}{c}{\text{Species 2}}\\
\hline\hline
&	&	G_2		&	S_2	\\
\hline
 \multirow{2}{*}{Species 1} & G_1 	& a_{G_1,G_2} &	a_{G_1,S_2} \\
&	S_1	&  a_{S_1,G_2} & a_{S_1,S_2} \\
 \hline\hline
\end{array}
\hspace{1cm}
\begin{array}{cccc}
& & \multicolumn{2}{c}{\text{Species 1}}\\
\hline\hline
&	&	G_1		&	S_1	\\
\hline
 \multirow{2}{*}{Species 2} & G_2 	& a_{G_2,G_1} &	a_{G_2,S_1} \\
&	S_2	&  a_{S_2,G_1} & a_{S_2,S_1} \nonumber \\
 \hline\hline
\end{array}
\end{equation}
where for example a generous member of species 1 obtains $a_{G_1,S_2}$ from an interaction with a selfish member of species 2, whereas the latter obtains $a_{S_2,G_1}$.
In our case, we have $a_{G_i,G_j} < a_{S_i,G_j}$ and $a_{S_i,S_j} < a_{G_i,S_j}$
for $i,j=1,2$.
This ordering of payoffs corresponds to a snowdrift game \cite{doebeli:EL:2005} (see Appendix) where there exists a point where the two strategies can coexist.
This is because if both the players are \textit{``Generous"} then one can get away with being \textit{``Selfish"}, but if both are \textit{``Selfish"} then it actually pays to be \textit{``Generous"} and chip in.
For a snowdrift game in a single species the coexistence point is stable, i.e. deviations from this point bring the system back to the equilibrium, because the deviators would always be disfavoured by selection.
However for a snowdrift game between two species, this coexistence point is unstable as each species would be better off being \textit{``Selfish"}, i.e.\ exploiting the deviation of the over species.

\begin{figure}[h]
\includegraphics[width=\columnwidth]{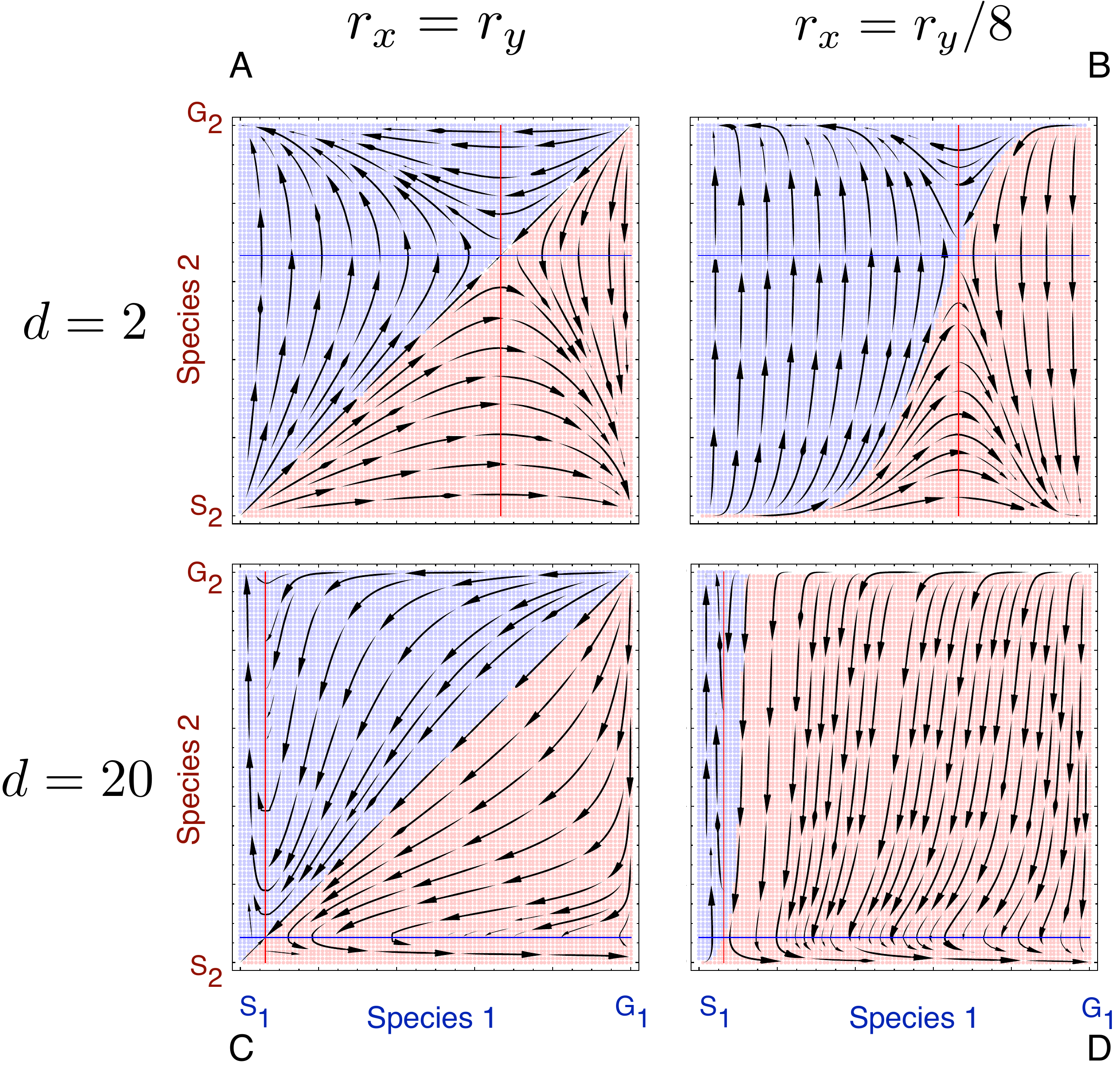}
\caption{
The composition of both species can range from all selfish $(S)$ to all generous $(G)$. 
If the other species is sufficiently generous, selfish behaviour is favoured in both species.
However, if the other species is selfish, generous behaviour is advantageous.  
This is captured by the snowdrift game discussed in the text.
For equal evolutionary rates, $r_x=r_y$, the basins of attraction for the two outcomes $(S_1, G_2)$ and $(G_1, S_2)$
are of equal size (Panels A and C).
The colours illustrate the regions leading to the outcomes favourable to species $1$ (blue shaded area leading to $(S_1, G_2)$) 
and species $2$ (red shaded area leading to $(G_1, S_2)$).
For a two player game, $d=2$, and $r_x=r_y/8$, the basin of attraction favourable to the slower evolving species $1$ grows substantially (Panel B) \cite{bergstrom:PNAS:2003}.
For a twenty player game, $d=20$, the basins of attractions have identical size for equal evolutionary rates, but the position of the internal equilibrium is shifted (Panel C).
When species $1$ evolves slower than species $2$ in this situation, most of the initial conditions lead to a solution which is unfavourable to species $1$ (Panel D).
Thus for 20 players instead of two, the Red King effect is reversed ($b=2$ and $c=1$).
\label{fig:compare}
}
\end{figure}

The frequency of players playing strategy \textit{``Generous"} ($G_1$) in species $1$ is given by $x$ and in species $2$ ($G_2$) by $y$.
The frequencies of players playing strategy \textit{``Selfish"} ($S_1$ and $S_2$) are given by $1-x$ and $1-y$ in species $1$ and $2$, respectively.
The fitness of the generous strategy in species 1, $f_{G_1}$ depends on the frequency $y$ of generous players in species 2, $f_{G_1}(y)$. Equivalently, the fitness of the generous strategy in species 2, $f_{G_2}$ depends on the frequency $x$ of generous players in species 1, $f_{G_2}(x)$. 
The replicator dynamics assumes that the change in frequency of a strategy is proportional to the difference between the fitness of that strategy and the average fitness of the species $\bar{f}$
 \cite{taylor:MB:1978,hofbauer:book:1998,hofbauer:BAMS:2003}. Thus, the time evolution of the frequencies of the \textit{``Generous"} players in the two species are 
\begin{eqnarray}
\dot{x} &= r_x x \left(f_{G_1}(y) -  \bar{f}_1(x,y) \right) \nonumber \\
\dot{y} &= r_y y \left(f_{G_2}(x) -  \bar{f}_2(x,y) \right).
\label{eq:orirepeqs}
\end{eqnarray}
The parameters $r_x$ and $r_y$ are the evolutionary rates of the two species.
We first recover the scenarios described in \cite{bergstrom:PNAS:2003}.
If the evolutionary rates are equal ($r_x=r_y$) and the evolutionary game is symmetric, then the basins of attraction of $(S_1, G_2)$ and $(G_1, S_2)$ are of equal size (Fig.\ \ref{fig:compare} Panel A).
For unequal evolutionary rates, the species which is evolving slower (in our case species $1$ with the rate $r_x=r_y/8$) has a larger basin of attraction (Fig. \ref{fig:compare} Panel B).
This asymmetry where most of the initial conditions lead to an outcome favouring the slower evolving species  has been termed as the \textit{Red King effect} \cite{bergstrom:PNAS:2003}.

We now extend the above approach to multiplayer games.
Extending the number of players from 2 to $d$ adds a polynomial nonlinearity to the fitness functions of the strategies (see Appendix).
For a multiplayer game we no longer have a $2 \times 2$ payoff matrix, but rather a payoff table. 
For this, we use the proposal from \cite{souza:JTB:2009} for a $d$-player snowdrift game, where the costs of being generous are divided among the \textit{``Generous"} players. 
In addition, only if there are at least $M$ \textit{``Generous"} players, a benefit is produced.
That is, for $k<M$ \textit{``Generous"} players, each one of them exhibits a loss of $c/M$ and the \textit{``Selfish"} players obtain nothing.
If there are at least $M$ \textit{``Generous"} players, then a benefit is produced. The \textit{``Generous"} obtain $b - c/k$ and the \textit{``Selfish"} obtain $b$ at no cost.
For species $2$ we can write down a different payoff setup which could have different values for $b$, $c$, $M$, $d$ etc.\ thus creating a ``bi-table" game.
For the time being, we assume that the payoff setups are symmetric for the two species and hence we just elucidate the details for species $1$.
The exact formulation of the payoffs and the calculations of fitness values are given in the Appendix.

Note that for $d=2$, $M=1$, $b=2$, and $c=1$ we recover the matrix used in \cite{bergstrom:PNAS:2003}.
Even for these new fitness functions for multiplayer games, the dynamics are still given by the replicator equations \cite{hauert:JTB:2006a,pacheco:PRSB:2009,gokhale:PNAS:2010}.
Also note that for two player games with $M=1$, there are four fixed points in which each species is either selfish or generous.
In addition, there is an internal fixed point given by $x = y = 2 (b-c)/(2 b - c) = \frac{2}{3}$.
The position and the stability of the fixed points is independent of the evolutionary rates, but as we will see, it depends on the number of players $d$.
For a $20$ player game ($d=20$), the basins of attractions are still of the same size, but the dynamics leading to the stable points on the vertices are completely different (Fig.~\ref{fig:compare} Panel C, $r_x=r_y$).
The internal equilibrium has now shifted to $x = y = 0.063$.
As before, we introduce an asymmetry in the evolutionary rates.
Interestingly, we find that for a $20$ player game (Fig.~\ref{fig:compare} Panel D, $r_x=r_y/8$) for the same asymmetric values of growth rates as in the two player case, most of the initial conditions lead to a stable point where species $2$ is selfish and species $1$ is generous ($G_1$, $S_2$).
This  reverses the result which we got from $d=2$.
Everything else being the same, in the presence of multiple players, the Red King effect is not observed in this example.

Next, we explore the process due to which the Red King effect vanishes.
The replicator solutions of the two species creates quadrants in the state space ($0 \leq x,y \leq 1$).
Of these quadrants, the top right and the bottom left are of special interest as they contain the curve that separates the two basins of attraction (the blue and red sections in Fig.~\ref{fig:compare}). 
Hence all points starting on one side of that curve lead to the same equilibrium.
Consider the top right quadrant.
Species $2$ is represented by the $y$-axis.
A faster evolution by species $2$ results in most of the initial conditions leading to the outcome favourable for species $2$, i.e. $(G_1,S_2)$.
An exactly opposite scenario is taking place in the bottom left quadrant.
Hence in this quadrant as species $1$ is evolving at a slower pace than species $2$, most of the initial condition here lead to an outcome favouring species $1$, i.e. $(S_1, G_2)$.
As long as the internal equilibrium is on the diagonal the Red King effect depends on the sizes of these quadrants.
Changing the number of players alters the sizes of these two influential quadrants.
For example consider the case of the twenty player game (Fig.~\ref{fig:compare} Panel C-D).
The size of the bottom left quadrant is reduced to such an extent that almost the whole state space leads to the outcome favourable for the faster evolving species.

The bottom left and the top right quadrants have equal size when the internal equilibrium is at $x = y = 0.5$.
For a fixed $b$ and $c$ we cannot select any arbitrary number of players $d$ to obtain this equilibrium, as $d$ is not a continuous variable.
If the equilibrium is above $x = 0.5$, then a decrease in the evolutionary rate can be beneficial as demonstrated by the Red King effect. 
Conversely, if the equilibrium is below $x=0.5$, then an increase in the evolutionary rate might be favourable.
Hence if the number of players positions the equilibrium at $x<0.5$ then the faster evolving species would be favoured.

Asymmetries have been considered in mutualistic species at the level of species or other properties of the system such as interactions \cite{noe:Ethology:1991}, interaction lengths \cite{johnstone:ECL:2002} or growth rates \cite{bergstrom:PNAS:2003}.
Asymmetry in the number of interacting partners has only been recently tackled \cite{wang:JRSI:2011}.
Going back to the example of ants and larvae, a single larva is tended to by multiple ants.
Thus while from each ant's point of view this is a two player game, for the larva this would be a multiplayer game.
Where such multiplayer games are feasible, it is also possible that a certain quorum needs to be fulfilled for the game to proceed.
Client fish have been shown to choose cleaning stations with two cleaners over solitary cleaners \cite{bshary:AB:2002}.
A certain number of ants are required to save a caterpillar from its predator.
It has been shown that the amount of secretions of a lycaenid larva is correlated to the number of attending ants \cite{axen:BE:1998}.
In the following two paragraphs we explore these two points, asymmetry in the number of players for the two species and different thresholds in either species to start off the benefits of mutualistic relations.

\begin{figure}
\includegraphics[width=\columnwidth]{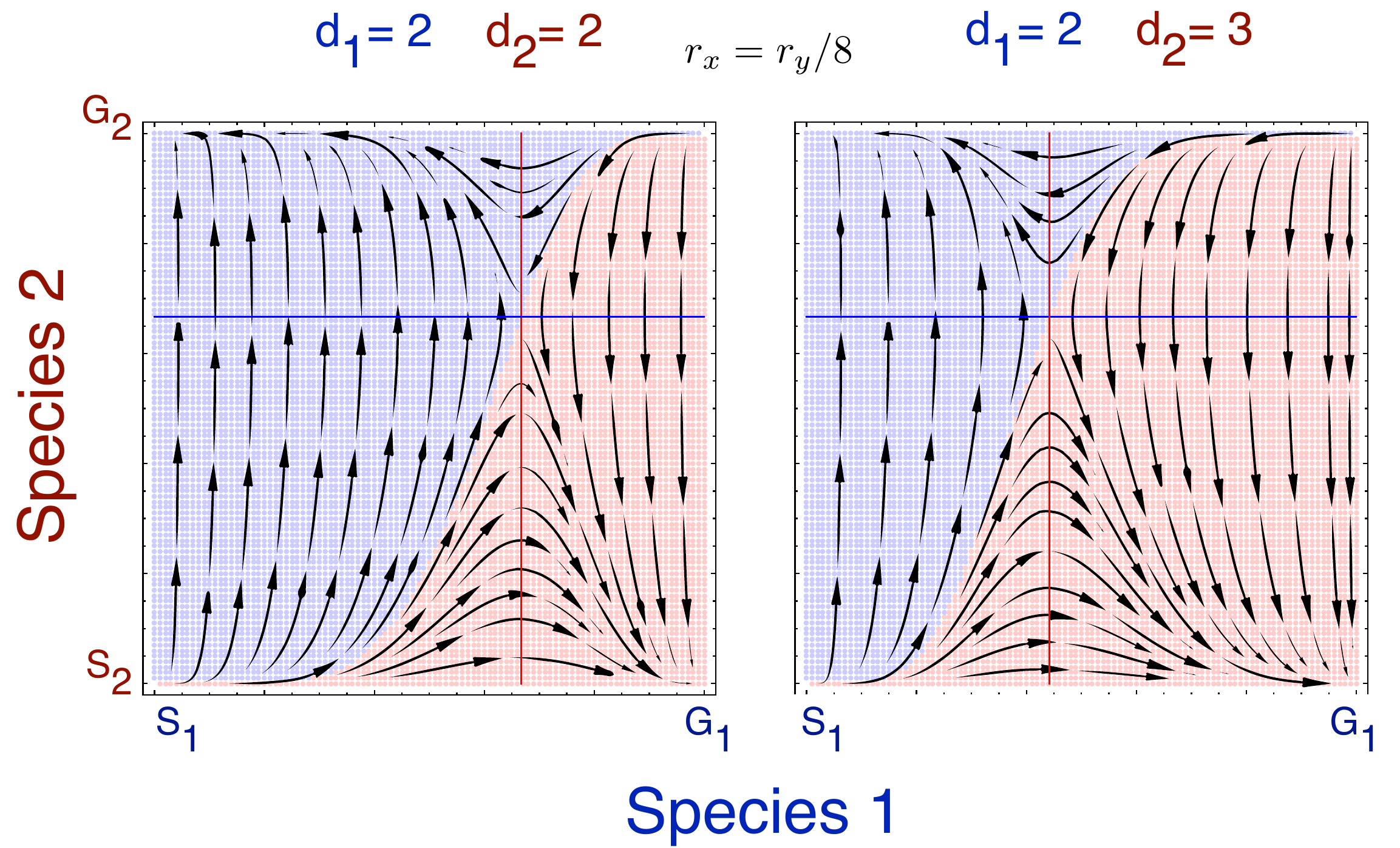}
\caption{
The Red King effect can be neutralised and/or even reversed if the number of players increases.
Here we show the scenario explored in \cite{bergstrom:PNAS:2003} on the left panel;
a two player game ($d_1 = d_2 = 2$) with $r_x=r_y/8$.
Most of the initial conditions lead towards the state favourable for species $1$, ($S_1,G_2$).
This changes when the number of players in species $2$ increases from $d_2=2$ to $d_2=3$,
i.e.\ now one individuals of species 2 interact with two individuals of species $1$. 
The horizontal and vertical lines denote the positions where the change in the strategy frequency is zero for species $1$ and $2$ respectively.
The solution for species $2$ (vertical line) moves towards smaller $x$, increasing the size of the top right quadrant.
For $d_1=2$ and $d_2=3$, it is given by $x = \frac{3 (2 b-c) - \sqrt{3 (4 b c - c^2)} }{2 (3 b -c)}$, whereas the 
the solution for species $2$ is still $ y = \frac{2(b-c)}{2 b - c}$ (see Appendix).  
Thus, the quadrant favouring the species with a faster evolutionary rate grows.
Since the number of players affects the size of these quadrants, it can eliminate or magnify the Red King effect.
}
\label{fig:counter}
\end{figure}

Instead of the single parameter $d$, now we have $d_1$ and $d_2$ as the number of players for the two species $1$ and $2$.
For symmetric evolutionary rates, if the two species play different player games ($d_1 \neq d_2$) then the basins of attraction become asymmetric.
Hence, if a species is currently at a disadvantage, a modification of the number of players or the evolutionary rate may put it on equal footing with the other species. 
Due to asymmetric number of players the sizes of the basins of attraction depends not just on the sizes of the quadrants but also on the shape of the curve separating the basins of attraction.
Thus, it is possible to counter the Red King effect by changing the number of interacting agents (Fig.\ \ref{fig:counter}).

Until now, we have considered that a single \textit{``Generous"} individual can generate the benefit of mutualism ($M=1$).
To begin with the simplest multiplayer case, we consider
a symmetric three player game with different thresholds in either species to start off the benefits of mutualistic relations (say for species $1$ the threshold is $M_1$ and similarly for species $2$ is $M_2$ where in general $M_i$ can range from $1$ to $d_i$).

The payoff matrices become asymmetric due to the different thresholds for the two species.
Here, it matters which dynamics we are studying, the usual replicator dynamics or the modified replicator equations (Fig.\ \ref{fig:thresholdsrep} and see Appendix) as they can result in different sizes of the basins of attraction.
The choice of dynamics depends on the details of the model system under consideration. 
Ultimately, the macroscopic dynamics can be derived from the underlying microscopic process \cite{traulsen:PRL:2005,black:TREE:2012}.
Manipulating the thresholds can also change the nature of the game from coexistence to coordination \cite{souza:JTB:2009},
which implies that different social dilemmas arise in multiplayer games.

\section{Discussion}

Interspecific relationships are exceedingly complex \cite{blaser:Nature:2007}.
The development of a game theoretical approach for such multiplayer mutualisms requires an approach beyond that arising typically in multiplayer social dilemmas \cite{bshary:ASB:2004}.
In a mutualistic framework, it is best for the two species to cooperate with each other.
We do not ask the question how these mutualisms arise. 
Rather when they do, what is the best strategy to contribute towards the common benefit \cite{bshary:book:2003,bowles:bookchapter:2003}?
It would be possible to include the interactions between the individuals of the same species, as has been explored experimentally recently \cite{wang:JRSI:2011}.
It has also been shown \cite{axen:BE:1998} that the amount of larval secretions is also influenced by the quality of the other larvae in the group.
But then, we would be shifting our focus from the problem of interspecific mutualism to intraspecific cooperation \cite{bshary:Nature:2008}.
Here, we have focused on the interspecific interactions, where the interacting partners are always picked from the other species \cite{schuster:BC:1981}.
Bergstrom and Lachmann have shown that in such a mutualistic scenario, the species which evolves slower can get away with being selfish and force the other species to make a generous contribution.
They termed this as the Red King effect.
If we include multiple players then the Red King effect is much more complex.

For simplicity, usually pairwise interactions are assumed in game theoretical arguments.
For modeling collective phenomena \cite{couzin:Nature:2005,sumpter:book:2010}, multiplayer games may be necessary.
The exact number of players is a matter of choice, though.
Group size distributions give us an idea about the mean group size of a species.
Instead of using pairwise interactions or an arbitrary number of individuals to form a group, we could use the mean group sizes as the number of interacting individuals.
Group size is known to be of importance in mutualisms \cite{wilson:AmNat:1983}.
As we have seen here, it can be a influential factor in deciding how the benefits are shared.
Countering the Red King or enhancing its effect is possible by altering the group size.
Hence can the group size itself be an evolving strategy?
The study of group size distributions has been tackled theoretically \cite{krause:book:2002,hauert:Science:2002, niwa:JTB:2003,hauert:PRSB:2006, hauert:BT:2008,veelen:JTB:2010,sumpter:book:2010,braennstroem:JMB:2011,pena:Evolution:2011} and empirically in various species ranging from house sparrows to humans \cite{zipf:book:1949,krause:book:2002,sumpter:book:2010,griesser:PLosOne:2011}.
In our example of ants and butterfly larvae, it has been observed that a larva was most successful in getting more ant attendants in a group of four larvae \cite{pierce:BES:1987}.
It would be interesting to see if the distributions in mutualistic species peak at the group size which is the best response to their symbiont partners choices.
This brings forth another of our assumption also implicit in \cite{bergstrom:PNAS:2003}.
The rate at which strategies evolve is assumed to be much faster than the rate at which the evolutionary rates or as just mentioned, the group size evolve.
If these traits are genetically determined then this assumption may no longer hold.
The rate of evolution is typically assumed to be constant, but it could well be a variable, subject to evolution.
We have seen that the number of players can affect whether evolving slower or faster is favourable.
It would then be interesting to determine the interplay between the evolving group size and the evolving evolutionary rate and what effect it has on the dynamics of strategy evolution.

Another method of introducing asymmetry is to have different payoff tables for the two species (i.e. different benefits and costs for the two species).
Also we have just considered two strategies per species.
Asymmetric number of strategies can induce further asymmetries in the interaction \cite{schuster:BC:1981}.
The intricacies of multiplayer games lend themselves to study such systems, but they also show that mutualistic interactions may be far more complex than often envisioned.
Applying multiplayer game theory to mutualism unravels this dynamics between species and can be used to understand the complexity of these non-linear systems.

\textbf{Acknowledgements}. We thank No\'emie Erin, Christian Hilbe, Aniek Ivens, Martin Kalbe, Michael Lachmann, Eric Miller and Istv\`an Scheuring  for helpful discussions and suggestions. 
We also thank the referees and the editor for their detailed and constructive input.
Financial support from the Emmy-Noether program of the DFG and from the Max Planck Society is gratefully acknowledged.

\begin{figure}[h]
\includegraphics[width=\columnwidth]{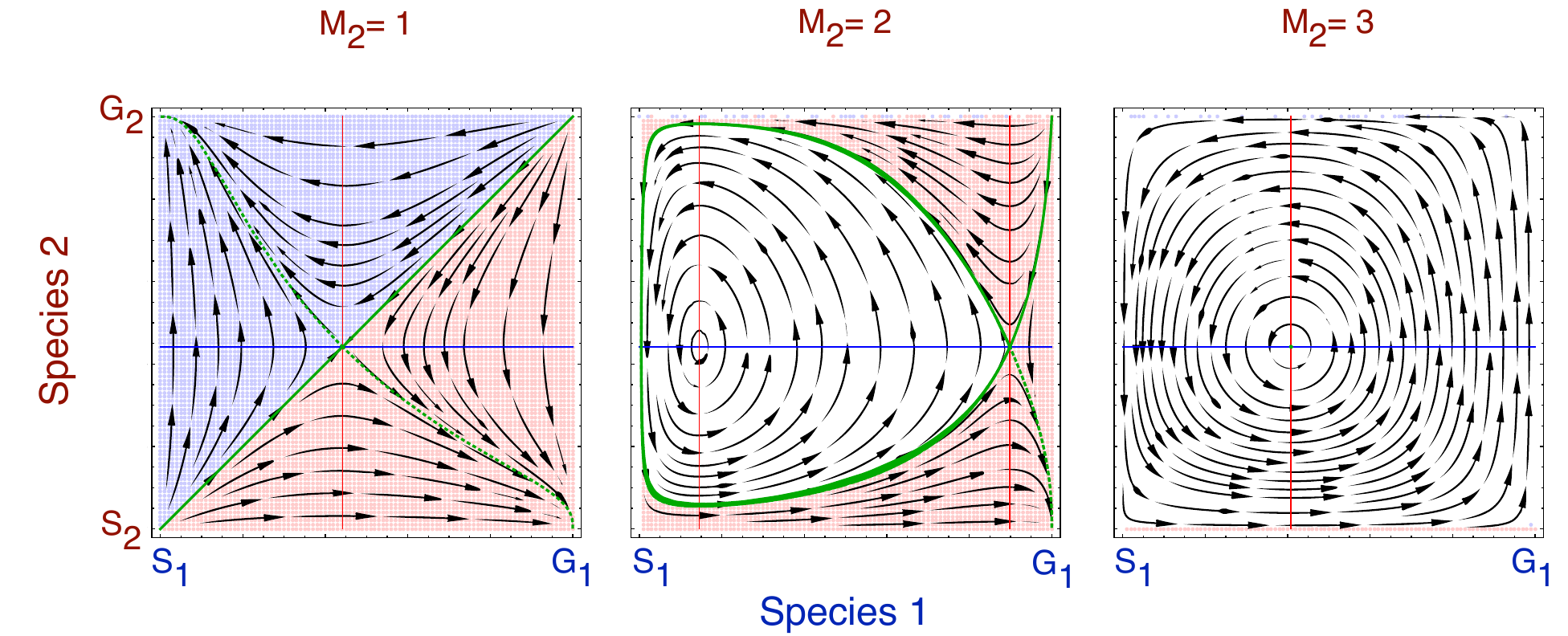}
\caption{
A three player game with asymmetric thresholds, but $r_x=r_y$.
Species $1$ and species $2$ are both playing a three player game.
For species $1$, it is enough if one individual is \textit{``Generous"} to produce the benefit ($M_1 = 1$).
For species $2$, however, the minimum number of \textit{``Generous"} players required to produce any benefit 
strongly affects the replicator dynamics. 
For $M_2=M_1=1$, we observe symmetric basins of attraction.
The manifolds for the saddle point plotted forward in time (dashed green lines) and backward in time (solid green lines) can be used to define the basins of attraction.
For $M_2=2$, we observe a region with closed orbits in the interior (white background), almost all initial conditions outside this region lead to ($G_1,S_2$). 
For $M_2=3$, we observe closed orbits in almost the whole state space.
To avoid negative payoffs and to facilitate the comparison with the modified dynamics (see Appendix), we have added a background fitness of  $1.0$ to all payoffs, but this does not alter the dynamics here.
}
\label{fig:thresholdsrep}
\end{figure}

\appendix

\renewcommand{\theequation}{A.\arabic{equation}}
\renewcommand{\thefigure}{A.\arabic{figure}}

\section{
Evolutionary games within a species and between two species
}
\subsection{Evolutionary games within a single species}
\label{withinsp}
In evolutionary game theory two player games with two strategies have been studied in great detail in infinitely large populations.
Typically, two players meet, interact and obtain a payoff. 
The payoff is then the basis for their reproductive success and hence for the change in the composition of the population \cite{maynard-smith:book:1982}.
Consider two strategies, $G$ and $S$.
We define the payoffs by $a_{i,j}$ where the strategy of the focal individual is $i$ and strategy of the other interacting individual is $j$.
For example, when an $G$ strategist meets another person playing $G$ she gets $a_{G,G}$.
She gets $a_{G,S}$ when she meets a $S$ strategist.
This leads to the payoff matrix
\begin{equation}\label{eq:twobytwo}
\begin{array}{ccc}
\hline\hline
 &$G$	&	$S$\\
\hline
$G$ 	& a_{G,G} &	a_{G,S} 
 \\
 $S$ 	&  a_{S,G} &a_{S,S} \nonumber \\
 \hline\hline
\end{array}
\end{equation}
The average payoffs for the two strategies are given by, %
\begin{eqnarray}
f_{G} (x) &=& a_{G,G} x + a_{G,S} (1-x) \nonumber \\
f_{S} (x) &=& a_{S,G} x + a_{S,S} (1-x), 
\end{eqnarray}
where $x$ is the frequency of the players with strategy $G$ and $1-x$ the frequency of players with strategy $S$.
Typically these average payoffs are directly considered as the average fitnesses of the two strategies.
This framework is used for biological systems, where strategies spread by genetic reproduction and often also for social systems where strategies spread by imitation.
This concept is valid for interactions within a species where the evolutionary dynamics of strategies within the species is of interest.
But quite often, interactions are not between two players, but between whole groups of players. 
Quorum sensing,  group hunting or climate preservation represent such multiplayer examples.
Such multiplayer games have been analyzed in the context of the evolution of cooperation \cite{hardin:Science:1968,hauert:PRSB:1997,kollock:ARS:1998,rockenbach:Nature:2006,milinski:PNAS:2006,milinski:PNAS:2008}, but only recently in a  more general sense \cite{pacheco:PRSB:2009,kurokawa:PRSB:2009,souza:JTB:2009,gokhale:PNAS:2010}.
The complexity brought to the table by such multiplayer interactions can be illustrated by adding just one more player in the previous setup.
Now instead of a two player game if we have a three player game then the payoff table is,
\begin{equation}\label{eq:threepl}
\begin{array}{cccc}
\hline\hline
 &$GG$	&	$GS$		&	$SS$\\
\hline
$G$ 	& a_{G,G,G} &	a_{G,G,S} &	a_{G,S,S} 
 \\
 $S$ 	&  a_{S,G,G} &a_{S,G,S}  &a_{S,S,S} \nonumber \\
 \hline\hline
\end{array}
\end{equation}
The fitnesses of the two types in such a case are given by,
\begin{eqnarray}
f_{G} (x) &=& a_{G,G,G} x^2 + 2 a_{G,G,S} x (1-x) + a_{G,S,S} (1-x)^2 \nonumber \\
f_{S} (x) &=& a_{S,G,G} x^2 + 2 a_{S,G,S} x (1-x) + a_{S,S,S} (1-x)^2 
\end{eqnarray}
where the nonlinearity of order two comes from the fact that each focal individual interacts with $2$ other individuals.
Since we assume that the order of players does not matter we get a binomial coefficient of $2$ when playing with two individuals with different strategies.

In whichever way the fitness is calculated, the evolution of the frequency of strategy $G$ is given by the replicator equation,
\begin{equation}
\dot{x} = x (f_{G} (x) - \bar{f}(x))
\end{equation}
where $\bar{f}(x) = x f_{G} + (1-x) f_{S}$ is the average fitness of the species.

\subsection{Evolutionary games between two species}
The above section was the traditional way of analyzing evolutionary games (either two or multiplayer) within a species but when species interact then we need to formalize the interactions in a different manner.
Just as in the previous section the average fitnesses were a function of the frequency of the strategies within the species (e.g. $f_{G}(x)$), here we assume that fitness depends on the frequency of the strategies in the other species (e.g. $f_{G}(y)$).
It is indeed possible to incorporate even more complexity by making the fitness depend on both the frequencies of strategies within the species and from the other species but that case is much more complicated and beyond the scope of this paper. 
Interested readers are referred to \cite{schuster:BC:1981c}.
For simplicity we stick to the assumption from coevolutionary models of interspecific dependence only \cite{roughgarden:TPB:1976,roughgarden:book:1983}.

\begin{figure}
\begin{center}
\includegraphics[scale=0.4]{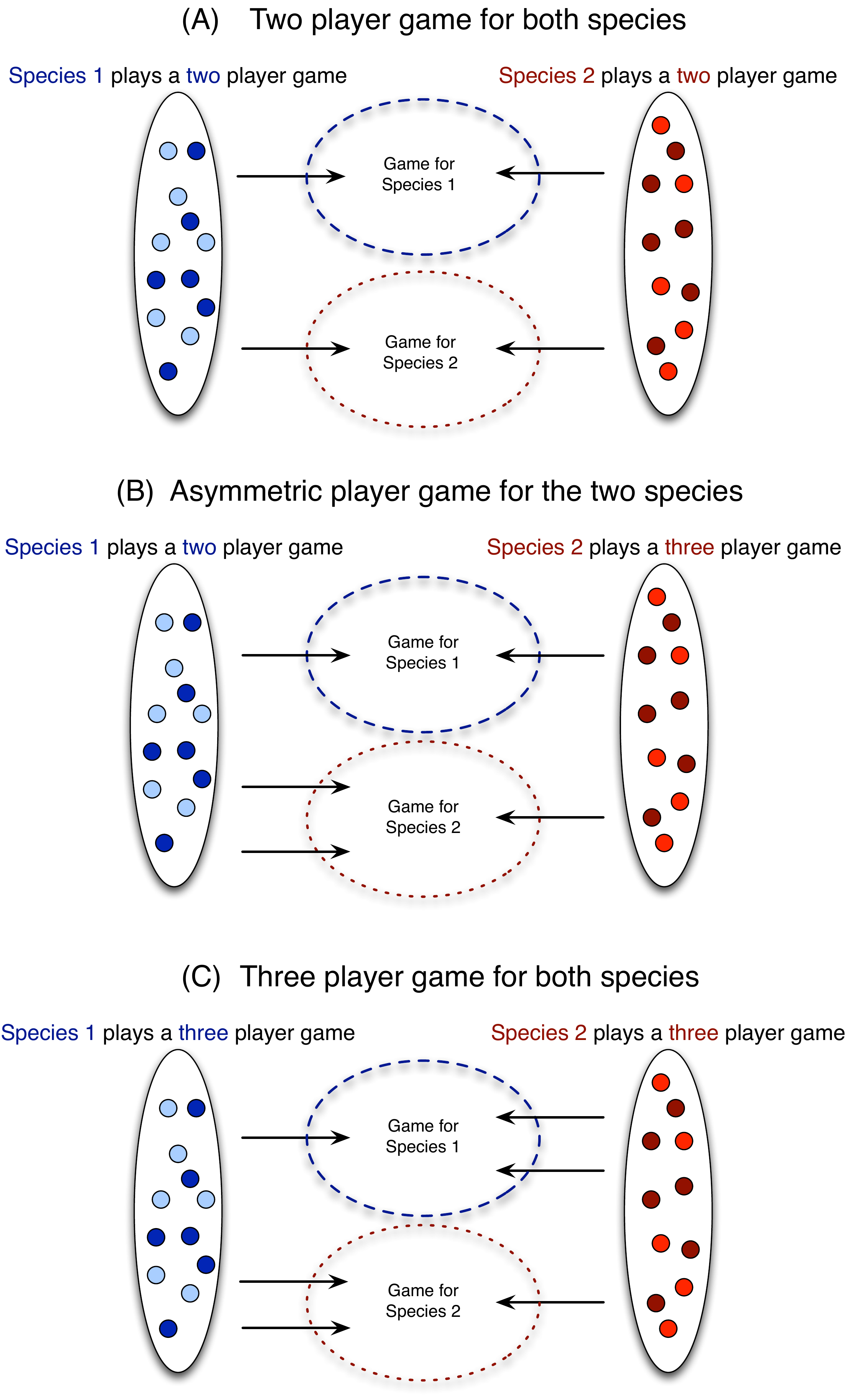}
\end{center}
\caption{
Evolutionary games between two species.
Each species is composed of players of two strategies denoted by the light and the dark circles.
(A) If both the species are playing a two player game then from both the species we choose one player each to form the interaction groups.
(B) Species $1$ plays a two player game, hence for this pairwise interaction we choose a single species two player to interact with a single species $1$ player (blue dashed oval). Species $2$ plays a three player game, hence one player of species $2$ interacts with two players of species $1$ (red dotted oval).
(C) If both species are playing a three player game then for a single player of species $1$ we pick two players from species $2$ (blue dashed oval) and vice versa for a single player of species $2$ (red dotted oval).
Thus in general for a $d_1$ player game for species $1$ we need to pick $d_1 - 1$ players from species $2$ and for a $d_2$ player game for species $2$ we need to pick $d_2 -1$ players from species $1$.
}
\label{fig:concept}
\end{figure}

For two player games between species (Fig. \ref{fig:concept} A)  now we need to pick one player from each species and they play the game.
For simplicity we assume that both the populations have the same strategies $G$ and $S$.
The fact that they belong to different species is now denoted by an index for $G_i$ and $S_i$ with $i = 1,2$.
Thus the payoff matrices describing the interactions
are,
\begin{equation}\label{}
\begin{array}{cccc}
& & \multicolumn{2}{c}{\text{Species 2}}\\
\hline\hline
&	&	G_2		&	S_2	\\
\hline
 \multirow{2}{*}{Species 1} & G_1 	& a_{G_1,G_2} &	a_{G_1,S_2} \\
&	S_1	&  a_{S_1,G_2} & a_{S_1,S_2} \\
 \hline\hline
\end{array}
\hspace{1cm}
\begin{array}{cccc}
& & \multicolumn{2}{c}{\text{Species 1}}\\
\hline\hline
&	&	G_1		&	S_1	\\
\hline
 \multirow{2}{*}{Species 2} & G_2 	& a_{G_2,G_1} &	a_{G_2,S_1} \\
&	S_2	&  a_{S_2,G_1} & a_{S_2,S_1} \nonumber \\
 \hline\hline
\end{array}
\end{equation}
with the respective average fitnesses now being,
\begin{eqnarray}
f_{G_1} (y) &=& a_{G_1,G_2} y + a_{G_1,S_2} (1-y) \nonumber \\
f_{S_1} (y) &=& a_{S_1,G_2} y + a_{S_1,S_2} (1-y)  \\
f_{G_2} (x) &=& a_{G_2,G_1} x + a_{G_2,S_1} (1-x)\nonumber \\
f_{S_2} (x) &=& a_{S_2,G_1} x + a_{S_2,S_1} (1-x).
\end{eqnarray}
Increasing the number of players in now an additional complication. 
Hence, we proceed by an addition of one more player in just one of the species to illustrate the intricacy.
Let us assume that species $1$ is playing a two player game while species $2$ plays a three player game (see Fig. \ref{fig:concept} B).
This means that we need to pick one player each from the two species to make up the two player game but we need two players from species $1$ and one player from species $2$ to make up the three player game.
This is because we have neglected intraspecific interactions as mentioned earlier.
If both the species play a three player game (see Fig. \ref{fig:concept} C) then we pick two player from each species and the games are again without including intraspecific interactions.
\begin{equation}\label{}
\begin{array}{ccccc}
& & \multicolumn{3}{c}{\text{Species 2}}\\
\hline\hline
& & G_2 G_2	&	G_2 S_2		&	S_2 S_2\\
\hline
 \multirow{2}{*}{Species 1} & G_1 	& a_{G_1,G_2,G_2} &	a_{G_1,G_2,S_2} &	a_{G_1,S_2,S_2} 
 \\
 & S_1 	&  a_{S_1,G_2,G_2} &a_{S_1,G_2,S_2}  &a_{S_1,S_2,S_2} \\
 \hline\hline
\end{array}
\hspace{0.5cm}
\begin{array}{ccccc}
& & \multicolumn{3}{c}{\text{Species 1}}\\
\hline\hline
& & G_1 G_1	&	G_1 S_1		&	S_1 S_1\\
\hline
 \multirow{2}{*}{Species 2} & G_2	& a_{G_2,G_1,G_1} &	a_{G_2,G_1,S_1} &	a_{G_2,S_1,S_1} 
 \\
 & S_2 	&  a_{S_2,G_1,G_1} &a_{S_2,G_1,S_1}  &a_{S_2,S_1,S_1} \nonumber \\
 \hline\hline
\end{array}
\end{equation}
and the respective average fitnesses are,
\begin{eqnarray}
f_{G_1} (y) &=& a_{G_1,G_2,G_2} y^2 + 2 a_{G_1,G_2,S_2} y (1-y) + a_{G_1,S_2,S_2} (1-y)^2  \nonumber \\
f_{S_1} (y) &=& a_{S_1,G_2,G_2} y^2 + 2 a_{S_1,G_2,S_2} y (1-y) + a_{S_1,S_2,S_2} (1-y)^2  \\
f_{G_2} (x) &=& a_{G_2,G_1,G_1} x^2 + 2 a_{G_2,G_1,S_1} x (1-x) + a_{G_2,S_1,S_1} (1-x)^2 \nonumber \\ 
f_{S_2} (x) &=& a_{S_2,G_1,G_1} x^2 + 2 a_{S_2,G_1,S_1} x (1-x) + a_{S_2,S_1,S_1} (1-x)^2.
\end{eqnarray}

Even if the fitness is for a two player game or for a multiplayer game, once it is calculated we can determine the change in the frequency of strategy $G_1$ and $G_2$ via the replicator equations, one for each species,
\begin{eqnarray}
\dot{x} &=& x \left(f_{G_1}(y) -  \bar{f}_1(x,y) \right) \nonumber \\
\dot{y} &=& y \left(f_{G_2}(x) -  \bar{f}_2(x,y) \right).
\end{eqnarray}
where $\bar{f}_i$ is the average fitness of species $i$.

\section{The snowdrift game}
\label{appA}
\subsection{Two player setting}
So far, we have described general games within and between species, now we turn to a particular game which of interest to us when considering mutualism.
The snowdrift game derives its name from the anecdote where two drivers are stuck in a snowdrift.
They must shovel away the snow (paying the cost $c$) to reach home (benefit $b$) but there are three possible outcomes to this scenario.
One of the driver shovels while the other stays warm in the car ($b-c$ and $b$), both the drivers share the workload and shovel away the snow ($b-c/2$ and $b-c/2$) or none of them gets out of the car and they both remain stuck ($0$ and $0$).

Putting this game in perspective of the two species (i.e. the two drivers represent the two different species) we get the matrix,\\
\begin{equation}\label{}
\begin{array}{cccc}
& & \multicolumn{2}{c}{\text{Species 2}}\\
\hline\hline
&	&	G_2		&	S_2	\\
\hline
 \multirow{2}{*}{Species 1} & G_1 	& b-c/2 &	b-c \\
&	S_1	&  b & 0 \\
 \hline\hline
\end{array}
\hspace{1cm}
\begin{array}{cccc}
& & \multicolumn{2}{c}{\text{Species 1}}\\
\hline\hline
&	&	G_1		&	S_1	\\
\hline
 \multirow{2}{*}{Species 2} & G_2 	& b-c/2 &	b-c \\
&	S_2	& b & 0 \nonumber \\
 \hline\hline
\end{array}
\end{equation}
where strategy $G$ stands for being \textit{``Generous"} and shoveling the snow while $S$ stands for being \textit{``Selfish"} and just sitting in the car.
For $b=2$ and $c=1$ we recover the matrix used in \cite{bergstrom:PNAS:2003}.

For the snowdrift game in a single population there exists a single, stable internal equilibrium.
Hence the population will evolve to a polymorphism which is a combination of \textit{``Generous"} and \textit{``Selfish"} players.
But in a two species system, this stable equilibrium turns into a saddle point, a small deviation from this fixed point can lead the system to one of the stable fixed point where one of the species is completely \textit{``Generous"} 
and the other one is completely \textit{``Selfish"}.

\subsection{Multiplayer setting}
\label{appB}

Following Souza et al. \cite{souza:JTB:2009},  
a multiplayer snowdrift game can be described by the payoff entries
\begin{eqnarray}
\Pi_{G_1} (k) &=& \begin{cases} b-\frac{c}{k} & \textrm{if } k \geq M \\  -\frac{c}{M} & \textrm{if } k < M \end{cases}
\\
\Pi_{S_1} (k) &=& \begin{cases} b & \textrm{if } k \geq M \\ 0 & \textrm{if } k < M. \end{cases}
\end{eqnarray}
In an interaction group, the \textit{``Selfish"} players get the benefit $b$ if the number of \textit{``Generous"} players, $k$, is greater than or equal to the threshold $M$.
For the \textit{``Generous"} individuals, their effort is subtracted from the payoffs.
The effort is shared if the quorum size is met ($\frac{c}{M}$), but is in vain for $k<M$.
For two player games we had $M=1$ but multiplayer games provide the possibility of exploring this threshold aspect of collective action games.
This relates back to the concept of quorum sensing \cite{sandoz:PNAS:2007,dyken:AmNat:2012} where a certain density has to be reached for the group benefit to be realized.
From these payoff entries we need to calculate the average fitnesses of the strategies.
For simplicity we just illustrate the fitnesses of the strategies in species $1$.
Just like in a three player game for species $1$ we needed to pick two individuals from species $2$, for a $d$ player game for species $1$ we need to pick $d-1$ other individuals from species $2$ (See Fig. \ref{fig:concept} C).
We assume that the groups are randomly assembled from the composition of species $2$.
Indeed due to this random sampling the composition of the formed groups is given by a binomial distribution.
Summing over all possible compositions of groups we arrive at  the average fitnesses of the two strategies in species $1$,
\begin{eqnarray}
f_{G_1} (y) &=& \sum_{i=0}^{d-1} \binom{d-1}{i}y^i (1-y)^{d-1-i} \Pi_{G_1}(i+1) \nonumber \\
f_{S_1} (y) &=& \sum_{i=0}^{d-1} \binom{d-1}{i}y^i (1-y)^{d-1-i} \Pi_{S_1}(i).
\label{fiteqs}
\end{eqnarray}
Following the same procedure for the two strategies in species $2$ leads to the average fitness of the two species
\begin{eqnarray}
\bar{f}_1 (x,y) &=& x f_{G_1} (y)+(1-x) f_ {S_1}(y)\nonumber \\
\bar{f}_2 (x,y) &=& y f_{G_2} (x)+(1-y) f_{S_2}(x).
\label{avgfiteqs}
\end{eqnarray}

\section{Dynamics in asymmetric conditions}

We have addressed two kinds of asymmetries in the game, the number of players and the thresholds in the two species.
We denote the number of players for species $1$ and species $2$ as $d_1$ and $d_2$, respectively.
That is if species $2$ is playing a $d_2$ player game it means that one player from species $2$ interacts with $d_2-1$ players of species $1$ (for e.g.	Fig \ref{fig:concept} B).
For an asymmetry in the thresholds we use the two parameters $M_1\geq1$ and $M_2\geq1$ for the two species, respectively.

\begin{figure}
\begin{center}
\includegraphics[width=\linewidth]{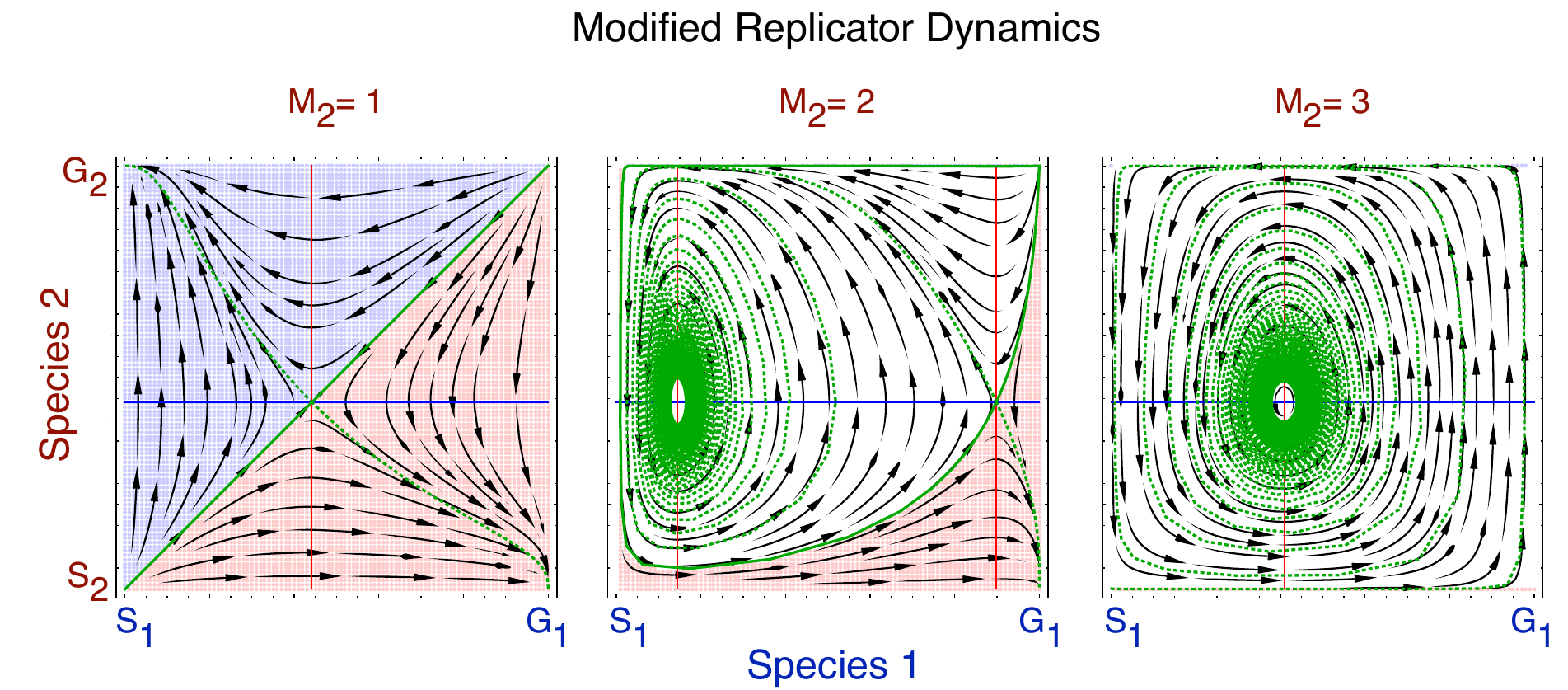}
\end{center}
\caption{
For the same parameters as in Fig.\ 3 (main text), the modified replicator dynamics given by Eqs.\ \ref{eq:repeqs} leads to a different dynamics.
For $M_1=M_2=1$ (left), the dynamics is the same as for the standard replicator dynamics. 
For $M_2=2$ (middle) and $M_2=3$ (right), the formerly neutrally stable fixed point in the interior becomes a stable focus. 
Moreover, for $M_2=2$, the basin of attraction of $(G_1,S_2)$ is much larger with the standard replicator dynamics. 
Again, we have added a background fitness of $1.0$ to all the payoff entries so that all payoffs are positive.}
\label{fig:thresholdsmodrep}
\end{figure}

For asymmetric bimatrix games, there is a difference in the dynamics between the standard replicator dynamics and the 
alternative dynamics put forth by Maynard Smith \cite{maynard-smith:book:1982}.
For this dynamics, the average fitness of each species appears as a denominator,
\begin{eqnarray}
\dot{x} &=& r_x x \left(f_{G_1}(y) -  \bar{f}_1(x,y) \right)/\bar{f}_1(x,y) \nonumber \\
\dot{y} &=& r_y y \left(f_{G_2}(x) -  \bar{f}_2(x,y) \right)/\bar{f}_2(x,y).
\label{eq:repeqs}
\end{eqnarray}
In our asymmetric bimatrix game, the fixed point stability is affected by the choice of the dynamics, in contrast to the case of symmetric games. 
In Fig.\ \ref{fig:thresholdsmodrep}, we illustrate that the dynamics is different between the usual 
replicator dynamics and Eqs. \ref{eq:repeqs}

For $d_1=d_2 \geq 5$, the exact coordinates of the fixed point must be computed numerically \cite{abel:AO:1824,stewart:book:2004}.

\end{document}